\documentclass{desyproc}

\begin{document}

\title{\vspace{-2.05cm}
\hfill{\small{DESY 11-178}}\\[1.27cm]
Dark Forces and Dark Matter in a Hidden Sector}

\author{{\slshape Sarah Andreas}\\[1ex]
Deutsches Elektronen-Synchrotron (DESY), Hamburg, Germany}

\contribID{andreas\_sarah}

\desyproc{DESY-PROC-2011-04}
\acronym{Patras 2011} 
\doi  

\maketitle

\begin{abstract}
Hidden sectors in connection with GeV-scale dark forces and
dark matter are not only a common feature of physics beyond the
Standard Model such as string theory and SUSY but are also
phenomenologically of great interest regarding recent
astrophysical observations. The hidden photon in particular is
also searched for and constrained by laboratory experiments,
the current status of which will be presented here.
Furthermore, several models of hidden sectors containing in
addition a dark matter particle will be examined regarding
their consistency with the dark matter relic abundance and
direct detection experiments.
\end{abstract}

\section{Motivation}
Hidden sectors are often predicted in string theories and
contained in various supersymmetric models as the source of
SUSY breaking. As the hidden sector (HS) is not charged under
the Standard model (SM) gauge groups and vice versa the two
sectors are not directly connected and only interact via
messenger particles. An example for a messenger is the hidden
photon $\gamma'$ which is a frequent feature of SM extensions
as extra hidden U(1) symmetries for example often remain in the
breaking of larger gauge groups or appear in string
compactifications.

On the other hand there are several observations in indirect
and direct dark matter detection experiments like PAMELA,
Fermi, DAMA and CoGeNT which favour dark matter (DM) models
with light messenger particles. Such a messenger, despite
mediating the DM scattering on nuclei, most importantly ensures
that the DM annihilation is at the same time leptophilic and
greatly enhanced by the Sommerfeld effect.

A GeV-scale mass for the hidden photon can be obtained quite
naturally both through the St\"{u}ckelberg and the Higgs
mechanism. The former, being the simplest mechanism to give
mass to abelian gauge bosons, can give for example in certain
string compactifications with D7-branes a mass to the hidden
photon according to $m_{\gamma'} \gtrsim
M_\mathrm{S}^2/M_\mathrm{Pl}$, which depends on the volume of
the extra dimension, i.e. the string-scale $M_\mathrm{S}$, and
the Planck scale
$M_\mathrm{Pl}$~\cite{Goodsell:2009xc,Cicoli:2011yh}. For
intermediate string-scales $M_S \sim 10^9-10^{10}$ GeV, which
are preferred by the axion decay constant and SUSY breaking
scales, this leads to $m_{\gamma'} \sim $ GeV-scale. In the
case of the Higgs mechanism where the symmetry breaking is
transferred by the kinetic mixing from the visible to the
hidden sector the hidden photon mass can be estimated by
$m_{\gamma'} \simeq \sqrt{g_Y g_h c_{2\beta}} \ v
\sqrt{\chi}$~\cite{Baumgart:2009tn,Morrissey:2009ur}. Assuming
that the kinetic mixing $\chi$ is generated supersymmetrically
at high scales, without light fields charged under both U(1)s,
it is of the order of a loop factor~\cite{Goodsell:2009xc} and
we impose the following relation with the hidden gauge coupling
$g_h$
\begin{equation}
\chi = \frac{g_Y g_h}{16 \pi^2} \ \kappa \label{eq-kappa}
\end{equation}
where $\kappa$ is an $\mathcal{O}(1)$ factor leading typically
to $\chi \sim 10^{-3}-10^{-4}$ and thus $m_{\gamma'} \sim $
GeV-scale.

\section{Hidden Photon: Constraints and future experiments}
\begin{wrapfigure}{r}{0.69\textwidth}
\vspace{-0.55cm} \centerline{
\includegraphics[width=0.34\textwidth]{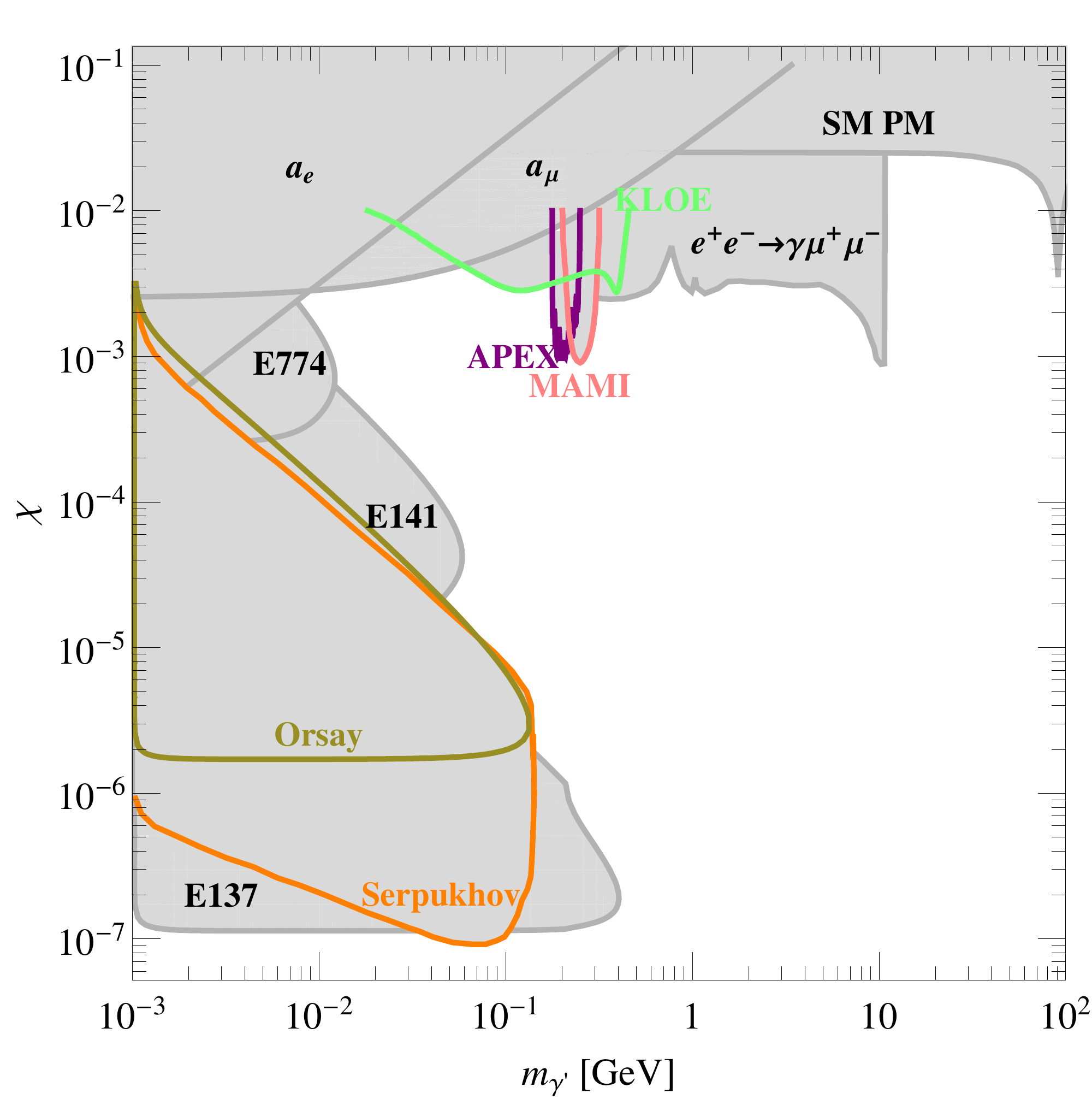}
\includegraphics[width=0.34\textwidth]{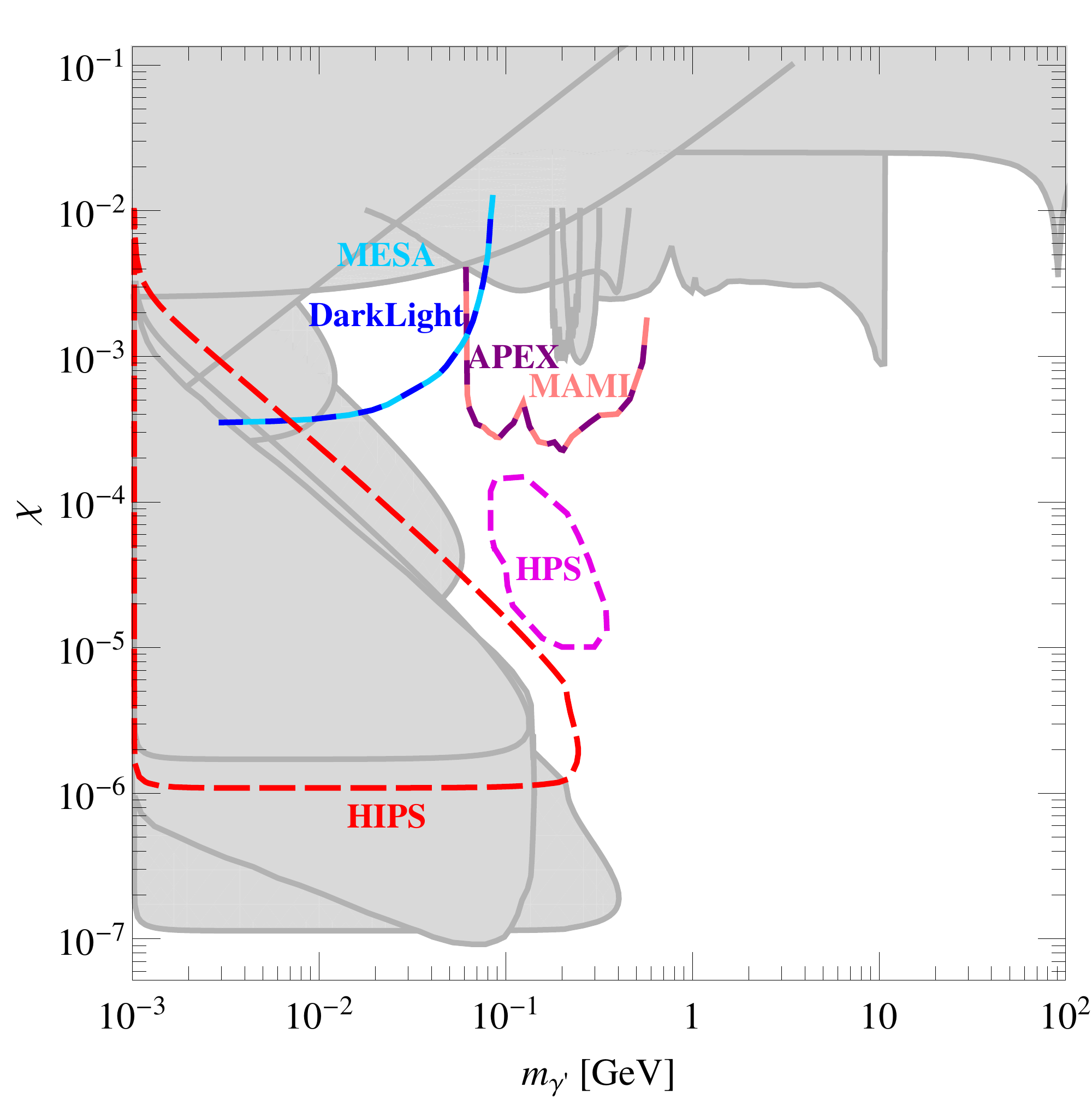}}
\vspace{-0.4cm} \caption{Hidden photon parameter space with
past and new exclusion regions \textit{(left)} as well as
projected sensitivities \textit{(right)}.} \label{fig-HPlimits}
\vspace{-0.35cm}
\end{wrapfigure}

We consider the most simple hidden sector containing only an
extra U(1) symmetry and the corresponding hidden photon
$\gamma'$ which kinetically mixes with the ordinary photon. The
most general Lagrangian for such a scenario, including a
mass-term $m_{\gamma'}$ for the hidden photon and the kinetic
mixing between $\gamma$ and $\gamma'$ parameterized by $\chi$,
is given by
\begin{equation}
{\cal L} = -\frac{1}{4}F_{\mu \nu}F^{\mu \nu} - \frac{1}{4}X_{\mu \nu}X^{\mu \nu}
- \frac{\chi}{2} X_{\mu \nu} F^{\mu \nu} + \frac{m_{\gamma^{\prime}}^2}{2}  X_{\mu}X^\mu + g_Y j^\mu_{\mathrm{em}} A_\mu,
\end{equation}
where $F_{\mu\nu}$ is the field strength of the ordinary
electromagnetic field $A_\mu$ and $X_{\mu\nu}$ the field
strength of the hidden U(1) field $X_\mu$. The hidden photon
can be constrained by and searched for in experiments
through its coupling to SM fermions which is possible due 
to the kinetic mixing with the photon. SM precision
measurements (SM PM)~\cite{Hook:2010tw}, the muon and electron
anomalous magnetic moment ($a_\mu, a_e$)~\cite{Pospelov:2008zw}
and a reinterpretation of the BaBar search for the
$\Upsilon(3S)$ decay to a pseudoscalar in the process $e^+ e^-
\to \gamma \mu^+ \mu^-$s~\cite{Essig:2010xa} exclude large
values of kinetic mixing as shown in grey in the left plot of
Fig.~\ref{fig-HPlimits}. Fixed-target experiments where
$\gamma'$ can be produced in
Bremsstrahlung off $e^-/p$-beams place additional constraints
from the non-observation of the decay $\gamma' \to e^+ e^-$.
Limits from past $e^-$-beam dump searches that have been
studied in~\cite{Bjorken:2009mm} are shown in the left plot of
Fig.~\ref{fig-HPlimits} as grey areas. The same graph also
contains as coloured lines the new limits set by currently
running (MAMI~\cite{Merkel:2011ze},
KLOE-2~\cite{Archilli:2011nh}, APEX~\cite{Abrahamyan:2011gv})
and reanalyzed older experiments
(Serpukhov~\cite{Blumlein:2011mv}, Orsay~\cite{Orsay}).
Experiments are closing in on the remaining region of the
parameter space and more is to be tested in the (near) future
by presently operating or planned fixed-target experiments at
JLab (APEX~\cite{Essig:2010xa}, HPS~\cite{talkSLAC-HPS},
DarkLight~\cite{Freytsis:2009bh,talkSLAC-DarkLight}), Mainz
(MAMI, MESA~\cite{talkJLab-MAMI}) and DESY
(HIPS~\cite{Andreas:2010tp}). The expected sensitivities of
those experiments are shown as coloured lines in the right plot
of Fig.~\ref{fig-HPlimits}.

\section{Hidden Dark Matter}
The results of this section are based on an analysis that has
been presented in detail in~\cite{Andreas:2011in}.
\vspace{-0.2cm}

\subsubsection*{Hidden Sector toy model: Dirac fermion dark matter}%
The simplest possible extension of the hidden sector studied in
the previous section is the addition of a Dirac fermion as dark
matter candidate
(cf.~\cite{Feldman:2006wd,Pospelov:2007mp,Chun:2010ve,Mambrini:2010dq}).
As the hidden photon mediates both the DM annihilation and the
DM scattering on nuclei it is essential for the determination
of the DM relic abundance and direct detection rate
respectively.
\begin{wrapfigure}{r}{0.34\textwidth}
\vspace{-0.2cm} \centerline{
\includegraphics[width=0.34\textwidth]{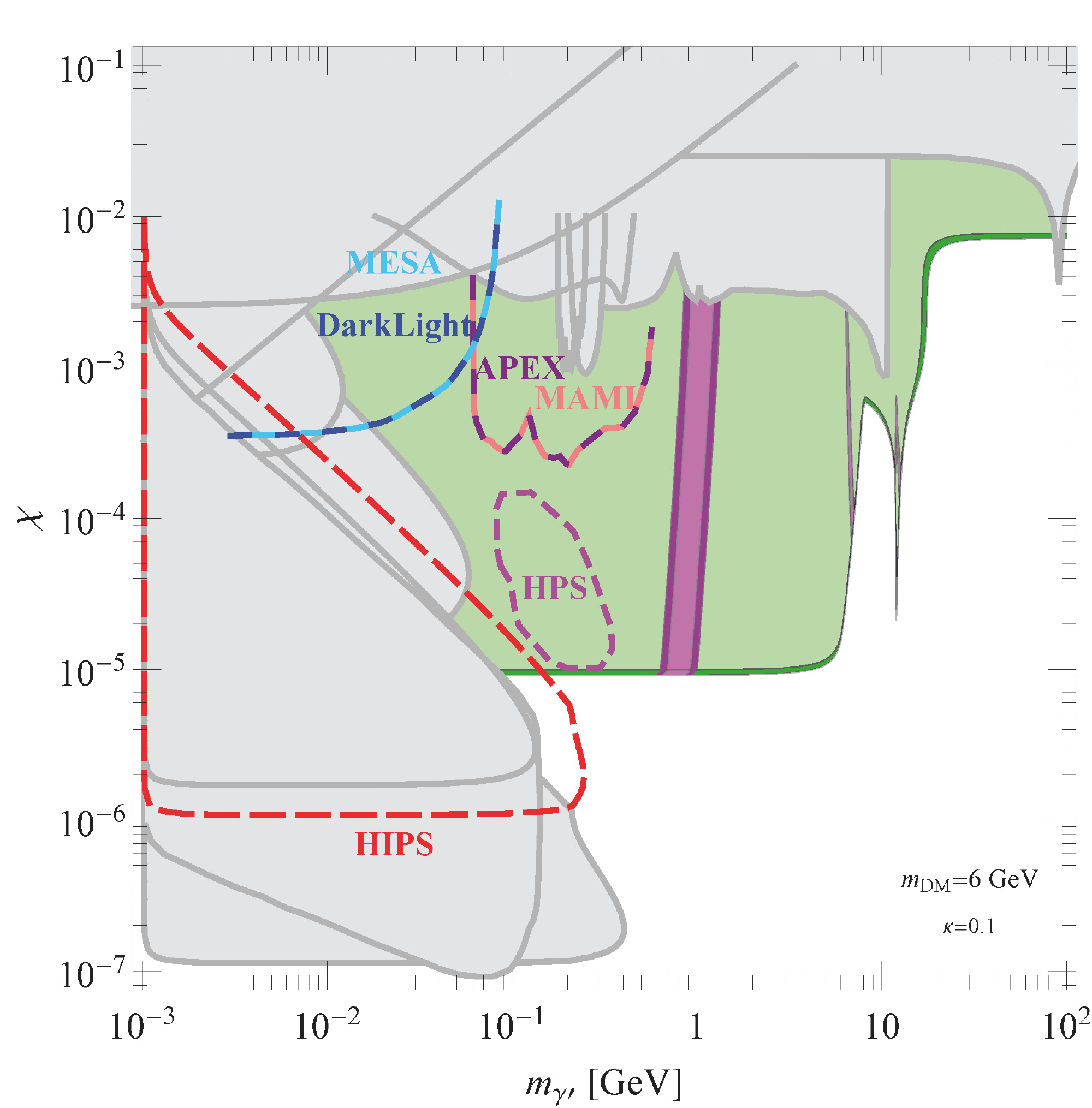}}
\vspace{-0.4cm} \caption{Toy model HS with relic abundance
(green) and CoGeNT region (purple) for the Dirac fermion DM.}
\label{fig-HDM-HPlimits} \vspace{-0.35cm}
\end{wrapfigure}
Using relation~(\ref{eq-kappa}) with $\kappa=0.1$ we find that
for a DM mass of 6 GeV the correct relic abundance can be
obtained on the dark green stripe in
Fig.~\ref{fig-HDM-HPlimits} while in the light green area the
contribution to the total DM density is only subdominant. The
spin-independent (SI) scattering on nuclei of the Dirac fermion
DM can explain the CoGeNT~\cite{Aalseth:2010vx} signal in the
purple band (90\% CL lighter 99\% CL darker purple). The
plotted cross section $\sigma_\mathrm{SI}$ has been rescaled by
the relic abundance for subdominant DM and fits the one found
in~\cite{Arina:2011si} to be compatible with CoGeNT for a
Standard Halo Model. Constraints from CDMS~\cite{Ahmed:2010wy}
and XENON~\cite{Aprile:2011hi} are not shown as they do not
apply to DM masses as light as 6 GeV. The excluded grey areas
and coloured lines for sensitivities of future experiments are
the same as in Fig.~\ref{fig-HPlimits}. A scan over the DM mass
allows to fill the complete parameter space as shown in
Fig.~\ref{fig-ScattPlot} on the left for $\kappa=1$ where dark
green corresponds to the correct relic abundance, light green
to subdominant and purple to CoGeNT allowed points (all points
are compatible with other direct detection limits). The effects
of varying the DM mass, the parameter $\kappa$ and the details
of the halo model have been studied in more detail
in~\cite{Andreas:2011in}.

\subsubsection*{Supersymmetric hidden sector: Majorana and Dirac fermion dark matter}
As a more sophisticated and better motivated model, we consider
three chiral superfields $S, H_+, H_-$ with $H_+$ and $H_-$
charged under the hidden U(1). Taking the superpotential
$W \supset \lambda_S S H_+ H_-$
and the dimensionless coupling~$\lambda_S$ this is the simplest
anomaly-free model possible without adding dimensionful
supersymmetric quantities. While we assume the MSSM in the
visible sector, the DM phenomenology of the HS depends on the
details of the SUSY breaking.
\begin{figure}[h!]
\vspace{-0.35cm} \centerline{
\includegraphics[width=0.33\textwidth]{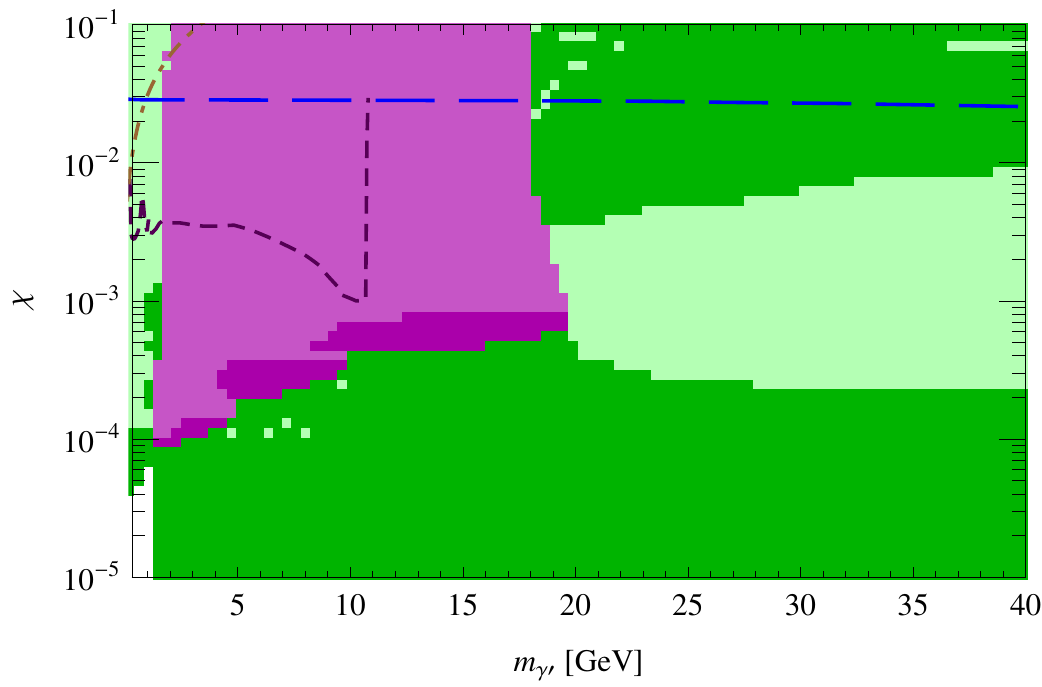}
\includegraphics[width=0.33\textwidth]{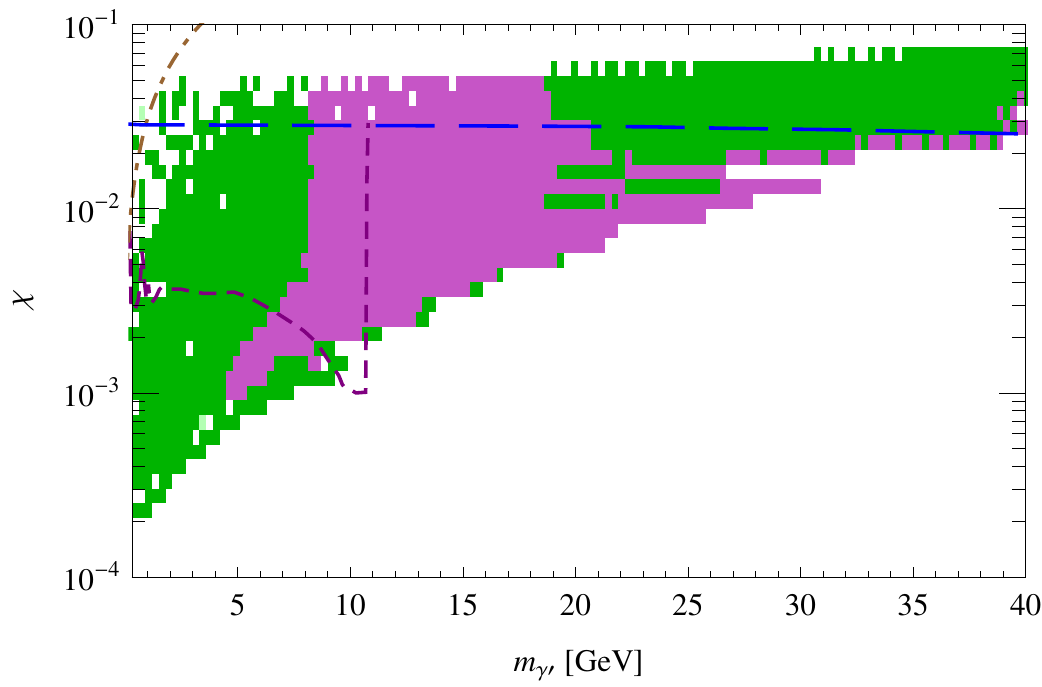}
\includegraphics[width=0.33\textwidth]{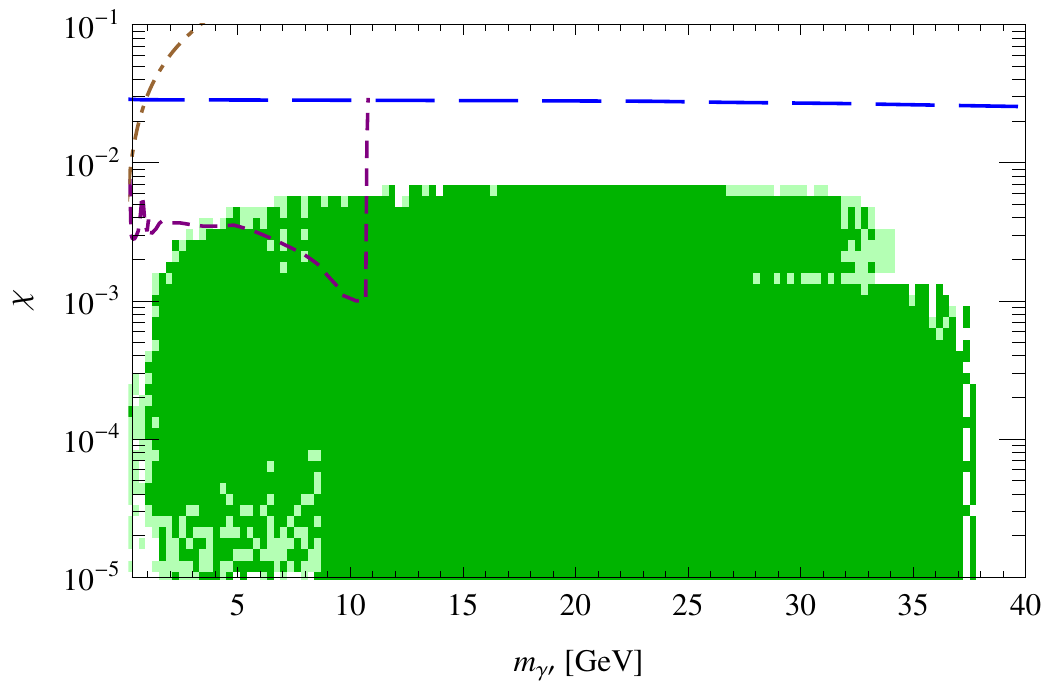}}
\vspace{-0.4cm} \caption{Scatter plots for the toy model \textit{(left)} and the SUSY HS with
visible sector induced breaking \textit{(middle)} as well as
radiative breaking domination \textit{(right)}. Lines are constraints from EWPT (long dashed),  BaBar (short dashed)
and muon $g$-2 (dash-dotted) similar to Fig.~\ref{fig-HPlimits}.} \label{fig-ScattPlot}
\vspace{-0.35cm}
\end{figure}
If the breaking of the hidden gauge symmetry is via the
effective Fayet-Iliopoulos term induced in the HS by the
kinetic mixing with the visible Higgs D-term we find that the
DM can be either a Dirac or a Majorana fermion. The former has
similar prospects for SI scattering as in the toy model as
shown in the middle plot of Fig.~\ref{fig-ScattPlot} where the
parameter $\kappa$ has been scanned over in the range
$0.1\leq\kappa\leq10$. The axial couplings of the latter lead
to dominantly spin-dependent (SD) scattering which is partly
constrained by experiments but without any chance of explaining
any of the signals in SI direct detection experiments. In the
case where the hidden gauge symmetry breaking is induced by the
Yukawa coupling $\lambda_S$, we find only a Majorana fermion as
lightest particle in the spectrum. As before this only shows in
SD direct detection experiments, hence the missing purple
colour for CoGeNT in the right plot of
Fig.~\ref{fig-ScattPlot}.

\section{Conclusions}
Hidden sectors are motivated by various aspects from top-down
(string theory, SUSY) and bottom-up (DM) and have a potentially
rich content like dark forces and dark matter which despite
their weak coupling can be phenomenologically interesting. We
reviewed past constraints on hidden photons as dark force and
showed the reach of future experiments. A simple toy model for
a HS with DM is found to be consistent with relic abundance and
direct detection and additionally provides the correct cross
section for explaining CoGeNT. The better-motivated
supersymmetric hidden sectors show some similarities with the
toy model but have a more complicated phenomenology where
spin-dependent scattering must also be taken into account.
Nevertheless, they also give viable models for DM with
interesting signatures in experiments.

\subsection*{Acknowledgments}
This work was done in collaboration with Mark Goodsell and
Andreas Ringwald.



\begin{footnotesize}

\end{footnotesize}


\end{document}